\documentclass[]{aa}  
\usepackage[pdftex]{graphicx}
\pdfoutput=1
\usepackage{txfonts}

\begin{document}

\title{High-resolution  Smoothed Particle Hydrodynamics simulations of
       the merger of binary white dwarfs}

\author{P. Lor\'en--Aguilar\inst{1,2}, 
        J. Isern\inst{3,2} \and 
        E. Garc\'{\i}a--Berro\inst{1,2}}
\institute{Departament de F\'\i sica Aplicada, 
           Escola Polit\'ecnica Superior de Castelldefels, 
           Universitat Polit\`ecnica de Catalunya,
           Avda. del  Canal Ol\'\i mpic 15, 
           08860 Castelldefels, Spain\
           \and       
           Institute for Space Studies of Catalonia,
           c/Gran Capit\`a 2--4, Edif. Nexus 104,   
           08034  Barcelona,  Spain\ 
           \and
           Institut de Ci\`encies de l'Espai, CSIC,  
           Campus UAB, Facultat de Ci\`encies, Torre C-5, 
           08193 Bellaterra, Spain\ }

\offprints{E. Garc\'\i a--Berro}

\date{\today}

\abstract{The coalescence of two white  dwarfs is the final outcome of
          a  sizeable fraction of  binary stellar  systems.  Moreover,
          this   process  has   been  proposed   to   explain  several
          interesting astrophysical phenomena.}
         {We  present   the  results  of  a   set  of  high-resolution
          simulations of the merging process of two white dwarfs.}
         {We  use an up-to-date  Smoothed Particle  Hydrodynamics code
          which  incorporates  very  detailed  input  physics  and  an
          improved  treatment   of  the  artificial   viscosity.   Our
          simulations have been done using a large number of particles
          ($\sim 4\times10^5$) and cover  the full range of masses and
          chemical  compositions of the  coalescing white  dwarfs.  We
          also compare  the  time evolution  of  the system during the 
          first phases of  the coalescence with  that obtained using a  
          simplified treatment of mass transfer, we  discuss in detail
          the characteristics of  the final  configuration, we  assess 
          the possible  observational  signatures  of the merger, like 
          the  associated  gravitational  waveforms  and  the fallback 
          X-ray  flares, and  we study  the long-term evolution of the 
          coalescence.}
         {The mass transfer rates  obtained during the first phases of
          the  merger   episode  are  in   agreement  the  theoretical
          expectations.   In  all   the  cases   studied   the  merged
          configuration is  a central  compact object surrounded  by a
          self-gravitating keplerian disk, except in the case in which
          two equal-mass white dwarfs coalesce.}
         {We find that  the overall evolution the system  and the main
          characteristics of the of  the final object are in agreement
          with  other previous  studies in  which  smaller resolutions
          were used. We also  find the the fallback X-ray luminosities
          are  of  the order  of  $10^{47}$  erg/s. The  gravitational
          waveforms are  characterized by the  sudden dissapearance of
          the signal in a few orbital periods.}


\titlerunning{High resolution SPH simulations of white dwarf mergers}

\authorrunning{P. Lor\'en--Aguilar et al.}

\maketitle


\section{Introduction}

The coalescence of two close white  dwarfs is thought to be one of the
most   common  endpoints   of   the  evolution   of  binary   systems.
Consequently, the  study of the coalescence process  is an interesting
issue, with  many potential applications.   Although the astrophysical
scenarios  in which the  coalescence of  two white  dwarfs in  a close
binary system can occur and  their relative frequencies have been well
studied --- see,  for instance, Yungelson et al.   (1994), Nelemans et
al.  (2001a,  2001b), and  the recent review  of Postnov  \& Yungelson
(2006) --- the  merging process  has received  little  attention until
recently.  The  pioneering works of Mochkovitch \&  Livio (1989, 1990)
who used an approximate method --- the so-called Self-Consistent-Field
method (Clement 1974) ---  and the full Smoothed Particle Hydrodynamic
(SPH) simulations  of Benz, Thielemann \& Hills  (1989), Benz, Cameron
\&  Bowers (1989),  Benz,  Hills  \& Thielemann  (1989),  Benz et  al.
(1990), Rasio \& Shapiro (1995) and Segretain, Chabrier \& Mochkovitch
(1997) were  the  only exceptions.   Most  of  these  early works  had
several  drawbacks.  For  instance, some  of  them did  not include  a
detailed  nuclear network or  the network  was very  simplistic, other
used a very small number  of SPH particles ($\sim 10^3$) and, finally,
other    did   not    discuss   the    properties   of    the   merger
configuration. Additionally,  all these early works  studied a reduced
set  of  masses  and  chemical  compositions and  used  the  classical
expression for  the artificial  viscosity (Monaghan \&  Gingold 1983).
This is an  important issue since it is well known  that SPH induces a
large  shear viscosity, which  is more  pronounced when  the classical
expression for the artificial viscosity is used.  Additionally, energy
dissipation  by  artificial viscosity  can  lead  to overheating  and,
hence,  the  peak temperatures  achieved  during  the  merger and  the
associated  nucleosynthesis depend  on  the choice  of the  artificial
viscosity.  However, the situation  has changed recently.  Guerrero et
al.   (2004)  opened  the  way  to  more  realistic  simulations.   In
particular,  in  these   calculations  the  standard  prescription  of
Monaghan \& Gingold  (1983) for the artificial viscosity  was used but
the  switch originally  suggested by  Balsara (1995)  was  employed to
partially  suppress the  excess  of dissipation.   More recently,  the
simulations of  Yoon et  al.  (2007) were  carried out using  a modern
prescription  for   the  artificial  viscosity   with  time  dependent
parameters (Morris  \& Monaghan 1997) which  guarantees that viscosity
is essentially absent  in those parts of the fluid in  which it is not
necessary, but there are  other prescriptions which are also suitable.
In particular the  prescription of Monaghan (1997), which  is based in
Riemann-solvers also yields excellent  results, and does not result in
overheating. This is the prescription we use in the present work. 

It is also interesting to realize that the number of particles used in
these  kind of  simulations  has increased  considerably  in the  last
years,  according  to the  available  computing  power. For  instance,
Segretain et  al.  (1997)  used $\sim 6\times  10^4$ SPH  particles to
simulate  the coalescence  of a  $0.9+0.6\, M_{\sun}$  system.  Later,
Guerrero et  al.  (2004)  studied a considerable  range of  masses and
chemical compositions of the merging white dwarfs employing a sizeable
number of  particles ($\sim 4\times  10^4$). This range of  masses and
chemical  compositions  included  6   runs  in  which  several  inital
configurations  were  studied,  involving  helium,  carbon-oxygen  and
oxygen-neon  white dwarfs.  Lor\'en-Aguilar  et al.   (2005) simulated
the  coalescence of  a system  of two  equal-mass  carbon-oxygen white
dwarfs  of   $0.6\,  M_{\sun}$.   However,  in  this   work  only  one
high-resolution simulation was done. More recently, Yoon et al. (2007)
have studied in detail the  coalescence of a binary system composed of
two white dwarfs  of masses $0.6$ and $0.9\,  M_{\sun}$ using $2\times
10^5$ SPH  particles.  However, only  one simulation was  presented in
this work.  It is thus clear that a thorough parametric study in which
several  white dwarf  masses  and chemical  compositions are  explored
using a large number of  SPH particles and a more elaborated treatment
of the artificial viscosity remains to be done.

Possible applications of these  kind of simulations include the double
degenerate  scenario  to  account  for  Type  Ia  supernova  outbursts
(Webbink 1984;  Iben \& Tutukov  1984) and the formation  of magnetars
(King,  Pringle \& Wickramasinghe  2001). Also  three hot  and massive
white dwarfs members  of the Galactic halo could be  the result of the
coalescence  of a  double white  dwarf binary  system (Schmidt  et al.
1992;  Segretain  et  al.   1997).   Additionally,  hydrogen-deficient
carbon and  R Corona  Borealis stars (Izzard  et al. 2007;  Clayton et
al. 2007) and extreme helium stars (Pandey et al. 2005) are thought to
be the  consequence of the merging  of two white  dwarfs. Finally, the
large metal  abundances found  around some hydrogen-rich  white dwarfs
with dusty  disks around them can be  explained by the merger  of a CO
and a  He white dwarf (Garc\'\i  a--Berro et al.  2007).  Last but not
least, the phase  previous to the coalescence of  a double white dwarf
close  binary  system  has been  shown  to  be  a powerful  source  of
gravitational  waves  which would  be  eventually  detectable by  LISA
(Lor\'en--Aguilar et al. 2005).

Depending  on  the  mass  ratio  of  both stars  and  on  the  initial
conditions of the binary system  the fate of double white dwarf binary
systems  is a  merging process  due to  the loss  of  angular momentum
through  gravitational  wave radiation.   Stars  orbit  each other  at
decreasing orbital  separations until  the less massive  one overfills
its Roche  lobe and  mass transfer begins.   According to  the initial
conditions mass transfer proceeds either in an stable or a dynamically
unstable regime. The stability of mass transfer is an important issue.
If the mass  transfer process is stable, mass  will flow at relatively
low  accretion rates  and the  whole  merging process  could last  for
several million years.   On the contrary, if mass  transfer proceds in
an unstable way, the whole  merging process finishes in a few minutes.
The  difference between the  two cases  relies on  the ability  of the
binary system to return enough  angular momentum back to the orbit. In
fact, there  are two competing processes.  On the one  hand, the donor
star is supported by the  pressure of degenerate electrons and, hence,
it  will expand  as it  loses mass,  thus enhancing  the mass-transfer
rate. On the other, if orbital angular momentum is conserved the orbit
will  expand  as   the  donor  star  loses  mass   thus  reducing  the
mass-transfer  rate.   The  precise  trade-off between  both  physical
processes  determines  the stability  of  mass  transfer. Guerrero  et
al. (2004) found that all the systems merged in a few hundred seconds,
corresponding  to mass  transfer  rates of  $\sim 10^{-2}\,  M_{\sun}$
s$^{-1}$.   Since the  Eddington rate  is of  the order  of $10^{-5}\,
M_{\sun}$ yr$^{-1}$  the most  massive white dwarf  cannot incorporate
the material of the disrupted secondary in such a short timescale and,
thus, the secondary forms a  hot atmosphere and a heavy keplerian disk
around the  primary. This  has been challenged  by the  simulations of
Motl et al.  (2002) and D'Souza  et al.  (2006).  These authors used a
grid-based three-dimensional finite-difference Eulerian hydrodynamical
code and  found that when the  stars are co-rotating  mass transfer is
stable.  Nevertheless, it should  be noted that these simulations were
done  using simplified  physical  inputs. For  instance,  they used  a
polytropic  equation of state.   More importantly,  grid-based methods
are  known to poorly  conserve angular  momentum. In  any case,  it is
clear  that to  assess the  stability of  mass transfer  large spatial
resolutions  are required  given the  degenerate nature  of  the donor
star,  since once  mass transfer  begins the  radius of  the secondary
increases very rapidly thereby increasing the mass-loss rate.

\begin{figure*}[t]
\includegraphics[width=225pt,angle=-90]{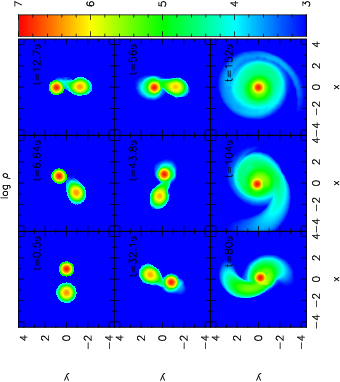}
\hspace{0.5cm} \includegraphics[width=225pt,angle=-90]{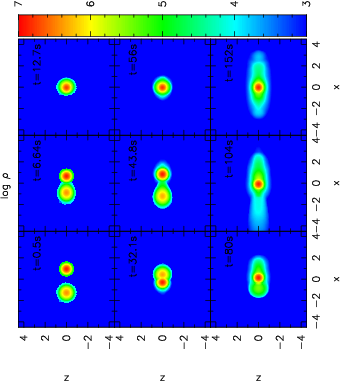}
\caption{Temporal evolution of the  density for the coalescence of the
         $0.6 + 0.8\, M_{\sun}$ double white dwarf binary system.  The
         positions of the particles  have been projected onto the $xy$
         plane (left panels) and in the $xz$ plane (right panels). The
         units of positions and densities are, respectively, $10^9$ cm
         and  $10^9$ g/cm$^3$.   Times are  shown in  the  right upper
         corner of each panel. These  figures have been done using the
         visualization  tool SPLASH (Price  2007). [Color  figure only
         available in the electronic version of the article].}
\label{snapshots-rho}
\end{figure*}

In the present  paper we study the coalescence  of binary white dwarfs
employing   an  enhanced  spatial   resolution  ($4\times   10^5$  SPH
particles) and  a formulation of  the artificial viscosity  which very
much reduces the excess of shear.  The number of particles used in our
simulations is  one order of magnitude  larger than those  used in our
previous simulations (Guerrero  et al. 2004) and within  a factor of 2
of   those  used   in   modern  simulations   (Yoon   et  al.    2007;
Lor\'en--Aguilar et al. 2005). This allows us to resolve smaller scale
lengths  --- by  a  factor $(10^5/10^4)^{1/3}\approx  2$  --- than  in
Guerrero et al.   (2004). Moreover, this is done for  a broad range of
initial  masses  and chemical  compositions  of  the coalescing  white
dwarfs,  in contrast  with most  modern  simulations in  which only  a
single  coalescence was  studied in  detail  (Yoon et  al. 2007).   In
particular we study the following  cases: $0.3 + 0.5\, M_{\sun}$, $0.4
+ 0.8\, M_{\sun}$, $0.6 +  0.6\, M_{\sun}$, $0.6 + 0.8\, M_{\sun}$ and
$0.6 + 1.2\,  M_{\sun}$.  Although we have computed  a large number of
mergers we will only discuss in  detail the results of the merger of a
$0.6 + 0.8  \, M_{\sun}$ binary system.  The main  results of the rest
of the simulations are only given in tabular form, but can be provided
upon request. Accordingly, we will devote most of the paper to compare
the results of this simulation with those previously available, and we
postpone the discussion of the  effects of the initial conditions to a
forthcoming  publication.   The paper  is  organized  as follows.   In
section 2 a  brief description of our SPH code is  given. In section 3
we describe our results and in section 4 we compare them with those of
other   authors.    In  section   5   we   discuss  our   simulations.
Specifically, we pay special attention to discuss the stability of the
mass  transfer episode.   This  is  done in  section  5.1, whereas  in
sections  5.2 and  5.3 the  possible observational  signatures arising
from the merging  process are studied. In particular,  we consider the
gravitational  wave  pattern  of  the  several  mergers  studied  here
(section 5.2) and  the X-ray emission that might  be expected from the
early phases of the disk evolution (section 5.3), while in section 5.4
the  long-term evolution  of  the merger  is  discussed.  Finally,  in
section  6  we summarize  our  major  findings,  we elaborate  on  the
possible implications of our work and we draw our conclusions.


\section{Input physics and method of calculation}

We  follow the  hydrodynamic evolution  of the  binary system  using a
Lagrangian  particle numerical code,  the so-called  Smoothed Particle
Hydrodynamics.   This method was  first proposed  by Lucy  (1977) and,
independently,  by Gingold  \&  Monaghan (1977).   The  fact that  the
method  is totally Lagrangian  and does  not require  a grid  makes it
specially  suitable for  studying  an intrinsically  three-dimensional
problem  like  the coalescence  of  two  white  dwarfs.  We  will  not
describe in  detail the  most basic equations  of our  numerical code,
since this is a well-known technique.  Instead, the reader is referred
to  Benz (1990)  where  the  basic numerical  scheme  for solving  the
hydrodynamic equations can be found, whereas a general introduction to
the  SPH method  can  be found  in  the excellent  review of  Monaghan
(2005).   However,  and  for  the  sake of  completeness,  we  shortly
describe the most relevant equations of our numerical code.

We use the standard polynomic  kernel of Monaghan \& Lattanzio (1985).
The gravitational forces are evaluated  using an octree (Barnes \& Hut
1986).  Our SPH code uses  a prescription for the artificial viscosity
based in  Riemann-solvers (Monaghan 1997).   Additionally, to suppress
artificial  viscosity forces  in  pure  shear flows  we  also use  the
viscosity switch of Balsara (1995).   In this way that the dissipative
terms are essentially  absent in most parts of the  fluid and are only
used where they  are really necessary to resolve  a shock, if present.
Within this  approach, the SPH  equations for the momentum  and energy
conservation read respectively

\begin{equation}
\frac{d\vec{v}_i}{dt} = - \sum_j m_j \left( \frac{P_i}{\rho_i^2} + 
\frac{P_j}{\rho_j^2} - \alpha \frac{v^{\rm sig}_{ij}}{\overline{\rho}_
{ij}}\vec{v}_{ij}\cdot \hat{e}_{ij} \right) \vec{r}_{ij} F_{ij}
\label{mom}
\end{equation}

\begin{equation}
\frac{du_i}{dt} = \frac{P_i}{\rho_i^2} \sum_j m_j \vec{v}_{ij} \cdot
\vec{r}_{ij} \overline{F}_{ij} - 
\frac{1}{2}\sum_j m_j \alpha \frac{v^{\rm sig}_{ij}}
{\overline{\rho}_{ij}}(\vec{v}_{ij} \cdot \hat{e}_{ij})^2 
|\vec{r}_{ij}| \overline{F}_{ij}
\label{uint}
\end{equation}

\noindent   where    $\overline{\rho}_{ij}=   (\rho_i+\rho_j)/2$   and
$\overline{F_  {ij}}\equiv  (F_i+F_j) /2  $,  and  $F$  is a  positive
definite function  which depends  only on $|\vec{r}|$  and and  on the
smoothing  kernel  $h$,  used   to  express  gradient  of  the  kernel
$\vec{\nabla}  W_{ij}  =  \overline{F}_{ij}\vec{r}_{ij}$.  The  signal
velocity is taken as  $v^{\rm sig}_{ij} = c_i+ c_j -4\vec{v}_{ij}\cdot
\hat{e}_{ij}$ and the rest of the symbols have their usual meaning. We
have found that $\alpha=0.5$ yields good results.

We  have found  that  it is  sometimes  advisable to  use a  different
formulation of  the equation of energy  conservation. Accordingly, for
each time step  we compute the variation of  the internal energy using
Eq.~(\ref{uint})  and   simultaneously  calculate  the   variation  of
temperature using:

\begin{equation}
\frac{dT_{i}}{dt}=-\sum_{j=1}^{N} \frac{m_j}{(C_v)_j}\frac{T_j}{\rho_i\rho_j}
\left[\left(\frac{\partial P}{\partial T}\right)_\rho\right]_j \vec{v}_{ij}
\cdot \vec{\nabla}_i W(|\vec{r}_{ij}|,h)+q_{\rm visc}
\label{temp}
\end{equation}

\noindent where  $q_{\rm visc}$  includes the contribution  of viscous
dissipation,  which  is  computed  in  a  way  analogous  to  that  of
Eq.~(\ref{uint}). For regions in which the temperatures are lower than
$6 \times  10^8$ K or the  densities are smaller than  $6 \times 10^3$
g/cm$^3$ Eq.~(\ref{uint}) is adopted, whereas Eq.~(\ref{temp}) is used
in the rest of the fluid.  Using this prescription we find that energy
is best conserved.
 
\begin{figure*}[t]
\includegraphics[width=225pt,angle=-90]{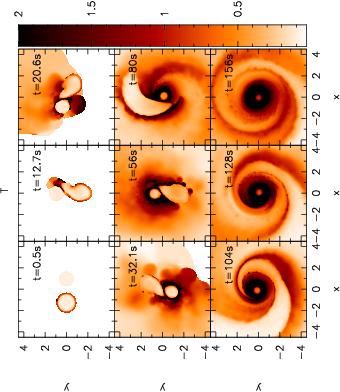}
\hspace{0.2cm} \includegraphics[width=225pt,angle=-90]{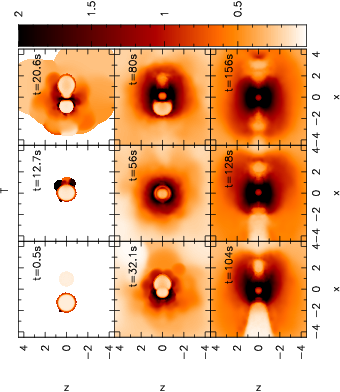}
\caption{Temporal evolution of the  temperature (in units of $10^9$ K)
         for  the  coalescence of  the  same  binary  system shown  in
         Fig.  \ref{snapshots-rho}.  The  positions  of the  particles
         have been projected onto the  $xy$ plane (left panels) and in
         the $xz$ plane (right  panels).  These figures have been done
         using  the  visualization tool  SPLASH  (Price 2007).  [Color
         figure  only  available  in  the electronic  version  of  the
         article].}
\label{snapshots-T}
\end{figure*}

\begin{table*}[t]
\centering
\begin{tabular}{cccccccccccccc}
\hline
\hline
\noalign{\smallskip}
 Run & $M_{\rm WD}$ 
     & $M_{\rm disk}$ 
     & $M_{\rm acc}$ 
     & $M_{\rm ej}$ 
     & $T_{\rm peak}$
     & $T_{\rm max}$
     & $R_{\rm disk}$
     & $H$
     & $\Delta t$
     & $E_{\rm nuc}$
     & $E_\nu$
     & $E_{\rm GW}$
     &$\omega$\\
\noalign{\smallskip}
\hline
\noalign{\smallskip}
0.3+0.5 & 0.62 & 0.18 & 0.12 & $10^{-3}$ & $6.0\times 10^8$    & $6.0\times 10^8$  &  0.2  & $6.1\times10^{-3}$ & 300 & $1\times10^{42}$ & $4\times10^{21}$ & $9\times10^{38}$ & 0.14\\
0.4+0.8 & 0.92 & 0.28 & 0.12 & $10^{-3}$ & $6.5\times 10^8$    & $6.0\times 10^8$  &  0.2  & $6.0\times10^{-3}$ & 166 & $1\times10^{44}$ & $5\times10^{24}$ & $4\times10^{39}$ & 0.23\\
0.6+0.6 & 1.10 & 0.10 & 0.50 & $10^{-3}$ & $6.3\times 10^8$    & $6.2\times 10^8$  &  0.07 & $5.5\times10^{-3}$ & 514 & 0                & $3\times10^{24}$ & $1\times10^{41}$ &0.27\\
0.6+0.8 & 1.10 & 0.30 & 0.30 & $10^{-3}$ & $1.6\times 10^9$    & $8.7\times 10^8$  &  0.2  & $5.0\times10^{-3}$ & 164 & $1\times10^{41}$ & $3\times10^{28}$ & $6\times10^{40}$ &0.33\\
0.6+1.2 & 1.50 & 0.30 & 0.30 & $10^{-3}$ & $1.0\times 10^{10}$ & $1.0\times 10^9$  &  0.2  & $4.4\times10^{-3}$ & 122 & $2\times10^{44}$ & $8\times10^{36}$ & $5\times10^{40}$ &0.52\\

\noalign{\smallskip}
\hline
\hline
\end{tabular}
\caption{Summary of  hydrodynamical results.  Masses and  radii are in
         solar  units, times  in seconds  and energies  in  ergs.  The
         maximum temperature achieved  during each simulation, $T_{\rm
         peak}$, and  the temperature of the hot  corona formed around
         the primary at the end of the simulations, $T_{\rm max}$, are
         discussed in  the text. The rotational  velocity is expressed
         in s$^{-1}$.}
\label{tab-hydro}
\end{table*}

The equation of state adopted for  the white dwarf is the sum of three
components.   The ions are  treated as  an ideal  gas but  taking into
account the Coulomb corrections (Segretain et al. 1994).  We have also
incorporated the pressure of photons,  which turns out to be important
when  the temperature  is high  and the  density is  small,  just when
nuclear  reactions  become   relevant.   Finally  the  most  important
contribution is the pressure  of degenerate electrons which is treated
integrating  the Fermi-Dirac integrals.   The nuclear  network adopted
here (Benz, Thielemann  \& Hills 1989) incorporates 14  nuclei: He, C,
O,  Ne, Mg,  Si, S,  Ar, Ca,  Ti, Cr,  Fe, Ni  and Zn.   The reactions
considered are captures of $\alpha$ particles, and the associated back
reactions, the fussion of two C nuclei, and the reaction between C and
O nuclei.  All the rates are taken from Rauscher \& Thielemann (2000).
The  screening factors  adopted  in this  work  are those  of Itoh  et
al. (1979).   The nuclear energy release is  computed independently of
the dynamical  evolution with  much smaller time-steps,  assuming that
the dynamical  variables do not  change much during  these time-steps.
Finally,  neutrino losses  have been  also included  according  to the
formulation of  Itoh et  al. (1996) for  the pair, photo,  plasma, and
bremsstrahlung neutrino processes.

Regarding  the   integration  method  we   use  a  predictor-corrector
numerical  scheme with  variable  time step  (Serna,  Alimi \&  Chieze
1996),  which  turns out  to  be  quite  accurate.  Each  particle  is
followed with  individual time steps.  With this  procedure the energy
and angular momentum  of the system are conserved  to a good accuracy.
To avoid numerical artifacts, we only use equal-mass SPH particles, as
it was  done in  Yoon et  al. (2007).  This  was not  the case  of the
simulations of Guerrero  et al. (2004) in which the  masses of the SPH
particles were different for each  one of the coalescing white dwarfs.
In order  to achieve an  equilibrium initial configuration  we relaxed
each  individual model star  separately, so  the two  coalescing white
dwarfs are spherically symmetric  at the beginning of our simulations,
as it  was the case in in  all previous simulations of  this kind.  In
all the cases  initially the two white dwarfs are  in a circular orbit
at a distance  larger than the corresponding Roche  lobe radius of the
less massive component.  The  systems are not synchronized because, at
least in the stage previous  to the coalescence itself, the time scale
for  loss of  angular momentum  due to  the emission  of gravitational
radiation is so small that it remains quite unlikely that there exists
any dissipation mechanism able to ensure synchronization (Segretain et
al.  1997).  This is the same initial configuration adopted by Yoon et
al.  (2007),  Guerrero et  al.  (2004) and  Segretain et  al.  (1997).
However, as  already stated, we  will study synchronized systems  in a
forthcoming  publication.   To  this   system  we  add  a  very  small
artificial acceleration  term which  decreases the separation  of both
components.  Once the secondary fills its Roche lobe this acceleration
term is supressed.  This procedure is quite similar to that adopted in
all  previous  works  (Guerrero  et  al.  2004;  Yoon  et  al.   2007;
Segretain et al.  (1997). We adopt this instant as our time origin.

The chemical compositions of the coalescing white dwarfs depend on the
mass  of each  star. White  dwarfs  with masses  smaller than  $0.45\,
M_{\sun}$ have pure He cores. For white dwarfs with masses within this
value  and  $1.1\,  M_{\sun}$  we  adopt  the  corresponding  chemical
composition, namely,  carbon and  oxygen, with mass  fractions $X_{\rm
C}=0.4$ and  $X_{\rm O}=0.6$ uniformingly distributed  through out the
core.  Finally,  white dwarfs more massive than  $1.1\, M_{\sun}$ have
ONe cores of  the appropriate composition (Ritossa et  al. 1996). This
is  essential in studying  the resulting  chemical composition  of the
merger, as it will be shown in  Sect. 3, and it is a clear improvement
with respect to recent  high-resolution simulations (Yoon et al. 2007)
in which  only a single system  of two carbon-oxygen  white dwarfs was
studied.


\section{Results}

\begin{table*}[t]
\centering
\begin{tabular}{lcccccccccccc}
\hline
\hline
\noalign{\smallskip}
\multicolumn{1}{c}{} &
\multicolumn{5}{c}{Disk} &
\multicolumn{5}{c}{Corona} \\
\noalign{\smallskip}
\hline
\noalign{\smallskip}
Run & 0.3+0.5 & 0.4+0.8 & 0.6+0.6 & 0.6+0.8 & 0.6+1.2 & 0.3+0.5 & 0.4+0.8 & 0.6+0.6 & 0.6+0.8 & 0.6+1.2 \\
\noalign{\smallskip}
\hline
\noalign{\smallskip}
He  & 0.94                & 0.98               &  0   &  0   &  $5\times10^{-5}$ & 0.72              & 0.68             & 0   & 0   & $5\times10^{-5}$  \\
C   & $2.4\times10^{-2}$  & $7\times10^{-3}$   &  0.4 &  0.4 &  0.39             & 0.11              & 0.12             & 0.4 & 0.4 & 0.26              \\
O   & $3.6\times10^{-2}$  & $9\times10^{-3}$   &  0.6 &  0.6 &  0.60             & 0.17              & 0.18             & 0.6 & 0.6 & 0.66              \\
Ne  & $9\times10^{-13}$   & $6\times10^{-10}$  &  0   &  0   &  $3\times10^{-3}$ & $3\times10^{-13}$ & $7\times10^{-7}$ & 0   & 0   & 0.07              \\
Mg  & $4\times10^{-14}$   & $5\times10^{-11}$  &  0   &  0   &  $7\times10^{-5}$ & $1\times10^{-12}$ & $1\times10^{-5}$ & 0   & 0   & $8\times10^{-5}$  \\
Si  & $1\times10^{-17}$   & $3\times10^{-14}$  &  0   &  0   &  $1\times10^{-5}$ & $1\times10^{-11}$ & $1\times10^{-4}$ & 0   & 0   & $5\times10^{-5}$  \\
S   & $2\times10^{-23}$   & $1\times10^{-19}$  &  0   &  0   &  $3\times10^{-7}$ & $1\times10^{-10}$ & $2\times10^{-4}$ & 0   & 0   & $5\times10^{-5}$  \\
Ar  & $<10^{-30}$         & $7\times10^{-27}$  &  0   &  0   &  $1\times10^{-7}$ & $5\times10^{-9}$  & $9\times10^{-4}$ & 0   & 0   & $4\times10^{-5}$  \\
Ca  & $<10^{-30}$         & $<10^{-30}$        &  0   &  0   &  $8\times10^{-7}$ & $1\times10^{-8}$  & $5\times10^{-4}$ & 0   & 0   & $1\times10^{-4}$  \\
Ti  & $<10^{-30}$         & $<10^{-30}$        &  0   &  0   &  $7\times10^{-7}$ & $2\times10^{-4}$  & $1\times10^{-2}$ & 0   & 0   & $1\times10^{-4}$  \\
Cr  & $<10^{-30}$         & $<10^{-30}$        &  0   &  0   &  $8\times10^{-7}$ & $4\times10^{-4}$  & $2\times10^{-3}$ & 0   & 0   & $2\times10^{-4}$  \\
Fe  & $<10^{-30}$         & $<10^{-30}$        &  0   &  0   &  $5\times10^{-6}$ & $2\times10^{-5}$  & $1\times10^{-5}$ & 0   & 0   & $6\times10^{-4}$  \\
Ni  & $<10^{-30}$         & $<10^{-30}$        &  0   &  0   &  $6\times10^{-4}$ & $2\times10^{-7}$  & $4\times10^{-8}$ & 0   & 0   & $1\times10^{-2}$  \\
Zn  & $<10^{-30}$         & $<10^{-30}$        &  0   &  0   &  $6\times10^{-6}$ & $2\times10^{-9}$  & $6\times10^{-10}$& 0   & 0   & $2\times10^{-5}$  \\

\noalign{\smallskip}
\hline
\hline
\end{tabular}
\caption{Averaged chemical composition (mass fractions) of  the  heavy
         rotationally-supported  disk and the  hot corona  obtained by
         the end of the coalescing process.}
\label{tab-chem}
\end{table*}

Fig.   \ref{snapshots-rho}   shows  the  temporal   evolution  of  the
logarithm  of the  density for  the coalescence  of the  $0.6  + 0.8\,
M_{\sun}$ white dwarf binary system.  In the left panels the positions
of the SPH particles have been projected onto the equatorial plane and
in the right panels onto the  polar plane.  Time (in seconds) is shown
on  the right  upper corner  of each  panel.  As  can be  seen  in the
uppermost left panels, the initial configurations of both white dwarfs
are symmetric.   Soon after,  the less massive  white dwarf  fills its
Roche lobe  and mass  tranfer begins,  as can be  seen in  top central
panel   of    this   figure.    The   top   right    panel   of   Fig.
\ref{snapshots-rho} shows that, after some time, the matter outflowing
the secondary hits the surface  of the primary white dwarf and spreads
on top  of it.   Note as well  that since  the radius of  white dwarfs
scales  as $\sim  M^{-1/3}$, as  the secondary  loses mass  its radius
increases and,  hence, the mass-loss rate of  the secondary increases,
thus leading to a positive  feedback of the process.  As a consequence
of this  positive feedback  an accretion arm  is formed  which extends
from the remnant of the  secondary white dwarf (central panels in Fig.
\ref{snapshots-rho}) to the surface  of the primary white dwarf.  This
accretion arm becomes entangled as a consequence of the orbital motion
of the coalescing white dwarfs  and adopts a spiral shape (bottom left
panel).  Ultimately,  the secondary is  totally disrupted and  a heavy
disk  is formed  around  the  primary (bottom  central  panel of  Fig.
\ref{snapshots-rho}).     The    bottom    right   panel    of    Fig.
\ref{snapshots-rho} shows that at time $t=152$ s the disk is still not
well  formed and  the  remnant of  a  spiral arm  still persists.   We
followed the evolution of this merger  for some more time and we found
that the  final configuration has  cylindrical symmetry, that  most of
the orbits of  the SPH particles belonging to  the secondary have been
circularized and that the  spiral pattern has totally disappeared.  At
the end of  the simulations the radial extension of  the disk is $\sim
0.2\,  R_{\sun}$,  whereas its  height  is  $\sim 5.0\times  10^{-3}\,
R_{\sun}$.

The temporal evolution  of the temperature for the merger  of a $0.6 +
0.8\, M_{\sun}$  binary system is shown in  Fig. \ref{snapshots-T}. As
can be  seen in  this figure  the material of  the secondary  is first
heated  by  tidal torques.  As  the  secondary  begins the  disruption
process this  material is  transferred to the  surface of  the primary
and,  consequently, it  is compressed  and its  temperature increases.
The peak temperatures ($T_{\rm peak}$) achieved during the coalescence
are  displayed in  Table  \ref{tab-hydro}  for each  one  of the  runs
presented in this paper.   For the $0.6+0.8\, M_{\sun}$ simulation the
peak temperature  is $T_{\rm  peak} \sim 1.6  \times 10^9$  K, clearly
larger than the carbon  ignition temperature $T_{\rm ign}\sim 10^9$ K,
and  occurs   during  the   first  and  most   violent  part   of  the
merger. However, a strong thermonuclear flash does not develop because
although  the temperature  in the  region  where the  material of  the
secondary first hits the primary increases very rapidly, degeneracy is
rapidly lifted,  leading to  an expansion of  the material,  which, in
turn, quenches the thermonuclear flash.  This is in agreement with the
results  of Guerrero et  al.  (2004)  and Yoon  et al.   (2007). Thus,
since these  high temperatures are  attained only during a  very short
time  interval   thermonuclear  processing  is  very   mild  for  this
simulation.   It is  also interesting  to compare  the  equatorial and
polar  distribution of  temperatures shown  in the  central  panels of
Fig.  \ref{snapshots-T}.   This  comparison  reveals that  the  heated
material is rapidly  redistributed on the surface of  the primary and,
as a  consequence, a hot corona  forms around the  primary. The spiral
structure previously  described can be more easily  appreciated in the
bottom right panels of  figure \ref{snapshots-T}. In fact, this spiral
structure persists for some more time.

\begin{figure*}[t]
\includegraphics[width=250pt]{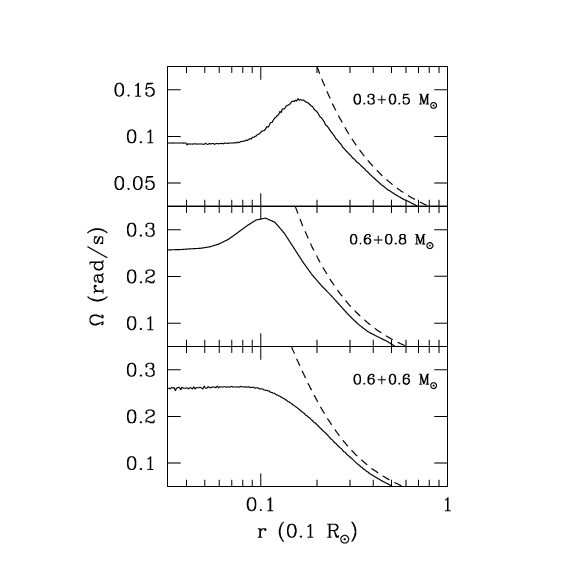}
\hspace{0.2cm} \includegraphics[width=250pt]{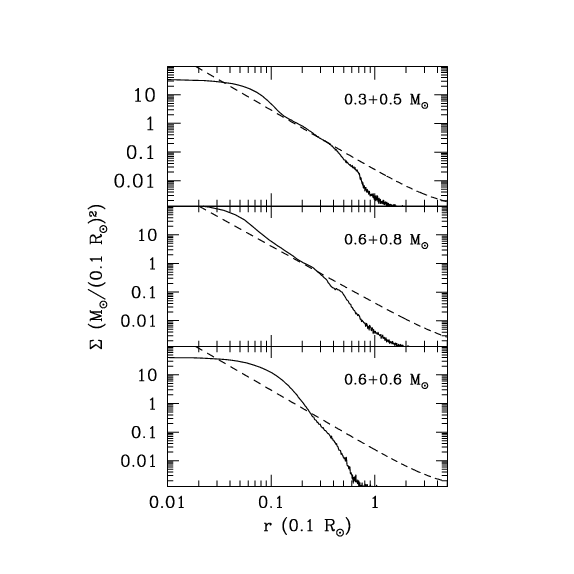}
\caption{Left panels: rotational velocity  of the merger products as a
         function  of  the radius.  For  the  sake  of comparison  the
         keplerian  velocity is also  shown as  a dashed  line.  Right
         panels:   surface   density   profiles  compared   with   the
         theoretical thin disk model profiles (dashed lines).}
\label{2Dvel}
\end{figure*}

In all the cases studied here a self-gravitating structure forms after
a  few  orbital  periods,  in  agreement with  our  previous  findings
(Guerrero et al. 2004) and with those of Yoon et al. (2007).  The time
necessary for its  formation depends on the system  being studied, and
ranges  from   $\sim  120$  seconds  to  $\sim   520$  seconds.   This
self-gravitating structure consists in all the cases but that in which
two equal-mass white dwarfs are  involved in a compact central object,
surrounded by  a heavy keplerian  disk of variable extension.   In the
case  in  which two  $0.6\,M_{\sun}$  white  dwarfs  are involved  the
configuration is rather different.  In  this last case the simmetry of
the systems  avoids the  formation of a  clear disk  structure, giving
rise instead to a rotating elipsoid around the central compact object,
surrounded by  a considerably smaller disk.   In table \ref{tab-hydro}
we summarize the  most relevant parameters of all  the mergers studied
here.  Columns two, three, four  and five list, respectively, the mass
of  the central  white  dwarf  obtained after  the  disruption of  the
secondary,  the mass  of  the  keplerian disk,  the  accreted and  the
ejected mass.  All the masses are expressed in solar units.  In column
six we show the peak temperatures achieved during the coalescence.  In
column seven we  display the temperature of the  hot corona around the
central object by the end  of our simulations, whereas in column eight
the  radius of the  disk is  shown. In  column nine  we list  the disk
half-thickness.  Column  ten displays the duration  of the coalescence
process and columns eleven, twelve and thirteen display the energetics
of  the  process.  In  particular  we  show  the thermonuclear  energy
released during the coalescence  process ($E_{\rm nuc}$), the neutrino
energy ($E_\nu$) and the energy  radiated in the form of gravitational
waves ($E_{\rm GW}$).  Finally, in column fourteen we list the angular
velocities of  the central compact remnants.  As can be  seen, for the
first two simulations the accreted  mass is approximately the same and
the same occurs  for the last two simulations.   As already commented,
the simulation  in which two  equal-mass white dwarfs are  involved is
rather special  and in this  case we do  not have properly  speaking a
disk, although a very flattened region with cylindrical symmetry forms
around a central object of  ellipsoidal shape. The mass of this region
is $\sim 0.1\, M_{\sun}$.  In all five cases the mass ejected from the
system  (those  particles which  acquire  velocities  larger than  the
escape velocity)  is very small ($\sim 10^{-3}\,  M_{\sun}$), and thus
the merging  process can be  considered as conservative.   The maximum
temperatures of the  coronae increase as the total  mass of the binary
system  increases.   It should  be  noted that  for  the  case of  the
$0.6+0.6\,M_{\sun}$  binary system the  maximum temperature  occurs at
the  center of  the merged  configuration.  We  have found  that these
temperatures are somewhat smaller  than those obtained in our previous
simulations (Guerrero et  al. 2004).  This is a  direct consequence of
the improved treatment of the artificial viscosity and of the enhanced
spatial  resolution.   For  instance,  we  have found  that  using  an
enhanced resolution and an improved prescription for the viscosity the
peak  temperature obtained  in the  simulation  in which  a $0.6$  and
$0.8\, M_{\sun}$ are involved  is $T_{\rm peak}\simeq 1.64\times 10^9$
K ---  see Table  1. When a  reduced number  of SPH particles  and the
classical  expression  for  the  artificial viscosity  are  used  this
temperature is $T_{\rm peak}\simeq 1.72 \times 10^9$ K, whereas when a
reduced number  of particles and an improved  artificial viscosity are
used the  peak temperature  turns out to  be $T_{\rm  peak}\simeq 1.68
\times 10^9$ K.   It is worth noting that the  radial extension of the
disks is  roughly the same for  all but one  the simulations presented
here  and  it  is considerably  smaller  for  the  case in  which  two
equal-mass  white dwarfs  are involved.   This is  a  natural behavior
since in this last case the central object is rather massive. Finally,
it is  as well interesting  to realize that  all the disks  are rather
thin, being the typical half-thickness of the order of $\sim 10^{-3}\,
R_{\sun}$, much smaller than  the typical disk radial extension, $\sim
0.2\, R_{\sun}$.

The chemical composition of the disk formed by the disrupted secondary
can be found for all the  simulations presented in this paper in Table
\ref{tab-chem}.   In this  table  we  show, for  each  of the  mergers
computed here,  the averaged chemical composition  (mass fractions) of
the  heavily rotationally-supported  disk  --- left  section of  table
\ref{tab-chem} --- and the hot  corona --- right section --- described
previously. For  the mergers in  which two carbon-oxygen  white dwarfs
are involved  the disk is mainly  formed by carbon and  oxygen and the
nuclear processing is  very small (see the peak  temperatures shown in
column ten  of Table  \ref{tab-hydro}). This is  not the case  for the
simulations in which  a lighter He white dwarf  is involved.  Since in
these cases  the Coulomb barrier is considerably  smaller, the shocked
material is nuclearly processed and heavy isotopes form.  This is more
evident  for  the   case  in  which  a  massive   He  white  dwarf  of
$0.4\,M_{\sun}$ is  disrupted by  a massive CO  white dwarf  of $0.8\,
M_{\sun}$  --- third and  eight columns  in Table  \ref{tab-chem}.  In
this case  the abundances in  the disk and  the hot corona  are rather
large. Note  as well that  the abundances of  heavy nuclei in  the hot
corona are much larger than those of the disk, indicating that most of
the nuclear reactions occur when the accretion stream hits the surface
of the primary.  Nevertheless, although  the disk is primarily made of
the He coming  from the disrupted secondary the abundances  of C and O
are sizeable and, moreover, the  disk is contaminated by heavy metals.
This has important consequences because it is thought that some of the
recently discovered metal-rich DA white dwarfs with dusty disks around
them  --- also  known as  DAZd  white dwarfs  --- could  be formed  by
accretion of a  minor planet.  The origin of  such minor planets still
remains  a mistery, since  asteroids sufficiently  close to  the white
dwarf  would  have not  survived  the  AGB  phase (Villaver  \&  Livio
2007). However, planet formation in these metal-rich disks is expected
to  be rather efficient,  thus providing  a natural  environment where
minor  planetary  bodies  could  be formed  and,  ultimately,  tidally
disrupted  to produce the  observed abundance  pattern in  these white
dwarfs (Garc\'\i  a--Berro et al.  2007).  Nuclear  reactions are also
important   in  the  case   in  which   a  regular   $0.6\,  M_{\sun}$
carbon-oxygen  white dwarf and  a massive  oxygen-neon white  dwarf of
$1.2\,  M_{\sun}$ are  involved.  In  this case  the  peak temperature
achieved  during the  coalescence is  rather high  $T_{\rm peak}\simeq
1.0\times 10^{10}$  --- see Table \ref{tab-hydro} ---  enough to power
carbon burning. Consequently, the chemical abundances of the keplerian
disk and  of the hot corona  are largely enhanced in  oxygen and neon,
which are the main products of carbon burning. We must add, however, a
cautionary remark  regarding the chemical compositions  of the mergers
studied here.   White dwarfs are  characterized in $\sim 80\%$  of the
cases by  a thin hydrogen  atmosphere of $\sim 10^{-4}\,  M_{\sun}$ on
top of a helium buffer of $\sim 10^{-2}\, M_{\sun}$.  In the remaining
$\sim  20\%$ of  the cases  the hydrogen  atmosphere is  absent. Small
amounts of  helium could indeed change the  nucleosynthetic pattern of
the  hot  corona  in  all  the  cases  studied  here.   Studying  this
possibility is beyond  the scope of this paper  and, thus, the changes
in the  abundances associated to burning  of the helium  buffer and of
the atmospheric hydrogen remain to be explored.

In  figure \ref{2Dvel}  we explore  the final  characteristics  of the
merged configuration.   We start discussing the left  panels of figure
\ref{2Dvel}  which show  the rotational  velocity of  the merger  as a
function of the  distance to the center of  the merged object. Clearly
in all  the cases  there is  a central region  which rotates  as rigid
solid --- see last column of table \ref{tab-hydro}.  This behavior was
already found in  Guerrero et al. (2004) and Yoon  et al.  (2007), and
it is a consequence of the conservation of angular momentum. On top of
this region a differentially  rotating layer is present.  This rapidly
rotating  region  is formed  by  material  coming  from the  disrupted
secondary, which has  been accumulated on top of  the primary and thus
carries the  original angular moment  of the secondary.   Finally, for
sufficiently large radius a rotationally-supported disk is found.  The
exact location where the disk begins can be easily found by looking at
the left panels of Fig.   \ref{2Dvel}, where the keplerian velocity is
also shown as  a dashed line.  The change in the  slope of the profile
of the rotational velocity clearly marks the outer edge of the compact
inner object and the beginnig of the disk.  All the disks extend up to
some solar radii  --- see column eight in  Table \ref{tab-hydro}.  The
stratification of surface densities of  these disks can be seen in the
left panels  of Fig.  \ref{2Dvel},  where we have plotted  the surface
density as a function of the  distance. For the sake of comparison the
theoretical surface  density of a  thin disk analytical  model (Livio,
Pringle \&  Wood 2005) is  also shown as  a dashed line.   Within this
model the  surface density of the  disk should be of  the form $\Sigma
\propto R^{-\beta}$.  We have used  $\beta=7/4$ to produce  the dashed
lines  in the right  panels of  Fig.  \ref{2Dvel},  very close  to the
value adopted by Livio et al.  (2005), $\beta=3/2$.  As can be seen in
this figure for the first two fiducial mergers studied here there is a
region where  the analytical  model and the  numerical results  are in
good agreement.   However, at large  enough distances the  SPH density
profile falls off more rapidly than that of the theoretical model.  In
the case of the merger of two equal-mass $0.6\, M_{\sun}$ white dwarfs
the  agreement is  poor.  In  this case,  the symmetry  of  the system
avoids the formation of a clear disk structure, giving rise instead to
a rotating ellipsoid around  the central compact object.  Moreover, it
can be shown that the angular momentum of the disk can be expressed in
terms  of the disk  radius $R_{\rm  disk}$ and  the disk  mass $M_{\rm
disk}$ as $J_z = \xi  M_{\rm disk} (G M_{\rm WD} R_{\rm disk})^{1/2}$,
where  $\xi=(2-\beta)/ (5/2  -\beta)= 1/3$.   The  theoretical angular
moments obtained using this equation  agree very well with the results
of our SPH simulations.

\begin{figure}[t]
\includegraphics[width=250pt]{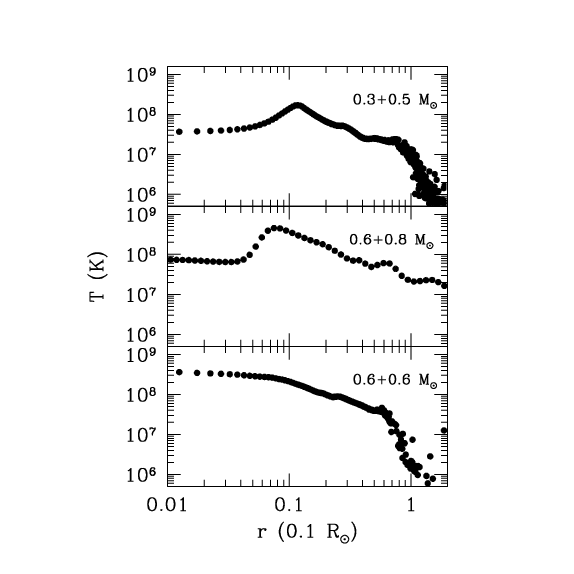}
\caption{Radially  averaged  temperature  profiles  as a  function  of
         radius.}
\label{2Dtemp}
\end{figure}

In Fig.  \ref{2Dtemp}  we show the temperature profiles  at the end of
the simulations for some of the mergers studied here. We have averaged
the temperatures of  those particles close to the  orbital plane.  The
average was done using cylindrical shells and the size of these shells
was chosen  in such a  way that each  of them contain  a significative
number of particles.  As can  be seen, for the $0.3+0.5\,M_{\sun}$ and
the  $0.4+0.8\,M_{\sun}$ systems, the  region of  maximum temperatures
occurs off-center, at the edge  of the original primary, in the region
of accreted and shocked material,  whereas for the merger in which two
equal-mass   $0.6\,M_{\sun}$  white   dwarfs   coalesce  the   maximum
temperature occurs at the center of the merged object, as it should be
expected.  These maximum temperatures are listed in the seventh column
of Table  \ref{tab-hydro}. In fact, the temperature  profiles shown in
this figure clearly show that the  cores of the primaries in the first
two  simulations remain  almost intact  and, hence,  are  rather cold.
These  cores,  in  turn,  are   surrounded  by  a  hot  envelope  wich
corresponds  to  the  shocked   material  coming  from  the  disrupted
secondary.  Nuclear reactions are responsible for the observed heating
of the  accreted matter, initially  triggered in the  shocked regions.
The  case  in  which  two  $0.6\,M_{\sun}$ white  dwarfs  coalesce  is
somewhat different.  In this case there is not a hot envelope around a
central object --- although a local maximum of temperatures can indeed
be appreciated at the edge  of the rapidly spinning central object, as
shown in the bottom panel  of Fig.  \ref{2Dtemp} --- and, instead, the
central region  of the compact  object is formed  by the cores  of the
merging white dwarfs.   Most of the temperature increase  in this case
is  due to  viscous  heating since  nuclear  reactions are  negligible
because the  increase in  temperature of the  shocked material  is not
enough to  ignite carbon.  In  all the cases  it is rather  apparent a
sizeable  dispersion of  temperatures in  the outermost  regions. This
dispersion is due  in part to the fact that  this region contains some
of the particles  that were ejected during the  first and most violent
phases of the merging process.

In summary,  we have  found that the  qualitative behavior of  all the
mergers  studied  here is  similar.  In  particular  the less  massive
component  of  the  binary  system  is  disrupted  in  a  few  orbital
periods.  Additionally,  we  find  that  nuclear  reactions  are  only
important in the  cases in which the secondary is a  He white dwarf or
in  the case  in which  the  primary (the  accretor) is  an ONe  white
dwarf.  The  total  ejected mass  is  very  small  in all  the  cases.
Finally, the  overall final configuration  is very similar in  all the
cases but in the case in which two equal-mass white dwarfs coalesce.


\section{Comparison with previous works}

As previously  discussed, all the  mergers studied here coalesce  in a
dynamical time  scale, regardless of its mass  ratio. Additionally, in
agreement  with  previous calculations,  the  central  regions of  the
remnant rotate as a rigid body. On top of this rapidly spinning core a
hot  corona forms  and, on  top  of it,  a heavy  keplerian disk.  The
question  is   now,  how  the  main  characteristics   of  the  merged
configuration compare  with those obtained in previous  works? Yoon et
al. computed  the coalescence  of a single  $0.6+0.9\,M_{\sun}$ binary
system and  the comparison is not straightforward,  but our $0.6+0.8\,
M_{\sun}$ run is  rather similar. The first thing to  be noted is that
the duration of the merger is very similar in both cases.  We obtain a
duration of  164 s  and Yoon et  al. (2007)  obtain $\sim 150$  s. The
central  density  of  the  rapidly  spinning core  is  in  both  cases
$\rho\sim 10^7$ g/cm$^3$.  The temperature of the core is $\log T=7.9$
whereas Yoon et al. (2007) obtain $\log T\sim 4.5$, but this is due to
our choice of the initial  temperature of the coalescing white dwarfs,
for which we adopted $T=10^7$ K.  However, the temperatures of the hot
coronae are remarkably similar in both cases, $\log T\sim 8.5$ and 8.6
respectively.  The  peak temperatures  attained during the  merger are
also  very similar  --- $1.7\times  10^9$  K and  $1.6\times 10^9$  K,
respectively.  However,   the  temperature  of  the   hot  coronae  is
considerably smaller in their  case $T_{\rm max}\simeq 5.8\times 10^8$
K.  This  value   has  to  be  compared  with   that  shown  in  table
\ref{tab-hydro},  $T_{\rm max}\simeq 8.7\times  10^8$ K.   However, it
should  be taken into  account that  Yoon et  al. (2007)  followed the
evolution  of the  merger for  much longer  times.  The  sizes  of the
resulting  disk are  also very  similar. Yoon  et al.  (2007) obtained
$8\times 10^9$  cm, whereas we obtain $1\times  10^{10}$ cm. Moreover,
despite  of   the  very   different  approaches  for   the  artificial
viscosities adopted  in the  work of  Yoon et al.   (2007) and  in the
present work, the rotational velocities of the central spinning object
are  very close,  0.21 s$^{-1}$  in the  case of  Yoon et  al. (2007),
whereas we obtain 0.26 s$^{-1}$.  However, we find that the rotational
velocities of  the hot coronae are somewhat  different. In particular,
Yoon  et al.  (2007) obtained  a $\omega\sim  0.54$ s$^{-1}$  while we
obtain 0.33 s$^{-1}$.  This could be due to the different treatment of
the artificial viscosity and to the different masses of the coalescing
white dwarfs.

The comparison with the results  of Guerrero et al. (2004) also yields
interesting results. For instance, the angular velocity of the central
compact object of the  $0.6+0.8\, M_{\sun}$ merger in the calculations
of Guerrero et al. (2004)  is 0.33 s$^{-1}$, somewhat larger than that
obtained  here.  Consequently,  the  central density  obtained in  the
simulations of Guerrero et  al. (2004) is smaller ($\rho\sim 6.3\times
10^6$  g/cm$^3$).   This is  due  to  our  improved treatment  of  the
artificial viscosity  which considerably  reduces the excess  of shear
and,  thus,  translates into  smaller  centrifugal  forces. Also,  the
position of the  hot corona is smaller in the  case of the simulations
presented here.   Specifically, the hot corona is  located at $0.008\,
R_{\sun}$ in  the simulations discussed  here, whereas in the  case of
Guerrero et  al. (2004) was  located at $0.01\, R_{\sun}$.   Also, the
temperature  of the  central object  is  smaller in  our case  ($T\sim
7.6\times 10^7$  K) than in  the case of  Guerrero et al.   (2004) for
which a central temperature of  $T\sim 3.8\times 10^7$ K was obtained,
even though  the initial temperatures  of the coalescing  white dwarfs
were the same in both  cases ($10^7$ K). Hence, our improved treatment
of the  artificial viscosity results  in a smaller overheating  and in
smaller shear.


\section{Discussion}

\begin{subsection}{Comparison with theory}

To obtain  a better  understanding of the  coalescence process  and to
compare  our  results  with  those  theoretically  expected,  we  have
numerically solved the equations of the evolution of the binary system
during  the mass  transfer phase.   The evolution  of a  binary system
during  this phase is  determined by  three basic  physical processes,
namely, gravitational wave emission,  tidal torques and mass transfer.
There is a  wealth of literature dealing with  this problem.  We adopt
as our starting point the analysis of Marsh et al. (2004) and the more
recent formulation  of Gokhale et  al.  (2007).  Within  this approach
the evolution of the orbital separation $a$ is given by

\begin{equation}
\frac{\dot{a}}{2a} = \frac{\dot{J}_{\rm GW}}{J_{\rm orb}} 
-\frac{I_1\left(\Omega-\omega_1\right)}{J_{\rm orb}\tau_{1}}-
 \frac{I_2\left(\Omega-\omega_2\right)}{J_{\rm orb}\tau_{2}}
-(q_{\rm a}-q)\frac{\dot{M}_2}{M_2}
\label{a}
\end{equation}

\noindent where $M_1$  and $M_2$ are, respectively, the  masses of the
accretor and of  the  donor, $q  \equiv  M_2/M_1$ is  the mass  ratio,
$\Omega$ is the orbital  velocity, $\omega_i$ are the spin velocities,
$I_i$   stands  for  the   moments  of   inertia,  $\tau_i$   are  the
synchronization timescales and $q_{\rm a}$ is defined as

\begin{equation}
q_{\rm a} \equiv 1+M_2\frac{j_2-j_1}{J_{\rm orb}}
\end{equation}

\noindent  being $j_1$  the specific  angular momentum  of  the matter
arriving to the accretor and  $j_2$ the angular momentum of the matter
leaving the donor.  In the calculations presented here we have adopted
for $j_1$ the expression for disk fed accretion:

\begin{equation}
j_1=\sqrt{GM_1R_1}
\end{equation}

\noindent whereas for $j_2$ we have

\begin{equation}
j_2\simeq  R^2_2\omega_2
\end{equation}

\noindent The first term in Eq. (\ref{a}) corresponds to the change in
the orbital separation due to  gravitational losses.  Due to the short
duration of the coalescing process, its contribution can be neglected.
The second and  third term describe the tidal  couplings. Finally, the
last  term  in Eq.   (\ref{a})  corresponds  to  the advected  angular
momentum.

The evolution of the Roche lobe radius $R_{\rm L}$ is given by

\begin{eqnarray}
\frac{\dot{R_{\rm L}}}{2R_{\rm L}} = \frac{\dot{J}_{\rm GW}}{J_{\rm orb}} 
&-&\frac{I_1\left(\Omega-\omega_1\right)}{J_{\rm orb}\tau_{1}}-
 \frac{I_2\left(\Omega-\omega_2\right)}{J_{\rm orb}\tau_{2}}\nonumber\\
&-&(q_{\rm a}-\frac{\zeta_{R_{\rm L}}}{2}-q)\frac{\dot{M}_2}{M_2}
\label{rl}
\end{eqnarray}

\noindent where $\zeta_{\rm R_{\rm  L}} \approx 0.30 +0.16q$, which is
valid for $0.01\leq  q \leq 1$ (Gokhale et al.  2007).  As can be seen
from this set  of equations, it s possible for a  system to change the
type of mass tranfer from stable to unstable or vice versa because $q$
has a dynamical  value.  Thus, to  determine  the exact  type of  mass
transfer of the  system it is necessary to study  how $q$ evolves with
time.

To solve  the previous equations, an expression  for the mass-transfer
rate is  needed.  The mass-transfer rate is  essentially determined by
the Roche lobe overfill factor,  wich is defined as $\Delta \equiv R_2
- R_{\rm L}$.   For the Roche lobe  radius we adopt  the expression of
Eggleton (1983).  In the case  of a $0.6\, M_{\sun}$ donor white dwarf
a polytropic equation of state  with $n=3/2$ can be adopted and, thus,
we have (Paczy\'nski \& Sienkiewicz 1972):

\begin{equation}
\dot{M}_2  = -\alpha W(\mu)\left(\frac{\Delta}{R_2}\right)^3
\label{delta}
\end{equation}

\noindent being

\begin{equation}
W(\mu)=\frac{\sqrt{\mu}\sqrt{1-\mu}}{(\sqrt{\mu}+\sqrt{1-\mu})^4}
\left(\frac{\mu}{R_1}\right)^3
\end{equation}

\noindent where $\mu=M_2/(M_1+M_2)$ and $\alpha$ is a smoothly varying
function of  the stellar parameters  which, following the  approach of
Gokhale et al.  (2007), we  have taken as a constant freely adjustable
parameter. The synchronization timescales have been computed using the
expressions of Campbell (1984)

\begin{eqnarray}
\nonumber
\tau_1 =\tau_1^0 \left(\frac{M_1}{M_2}\right)^2\left(\frac{a}{R_1}\right)^6 \\
\tau_2 =\tau_2^0 \left(\frac{M_2}{M_1}\right)^2\left(\frac{a}{R_2}\right)^6 
\end{eqnarray}

\noindent  where,  again,  the  normalization factors  $\tau_1^0$  and
$\tau_2^0$  are  freely  adjustable  parameter (Marsh  et  al.   2004;
Gokhale et  al.  2007).  With all  these inputs the  equations for the
evolution of the  binary system --- Eqs. (\ref{a})  and (\ref{rl}) ---
together  with the  equations for  the evolution  of the  spin angular
velocities of each of the components

\begin{equation}
\dot{\omega}_i = \left(\frac{j_i}{I_i}\right)\dot{M_i}-
\left(\frac{\dot{I_i}}{I_i}\right)\omega_i+\frac{\Omega-\omega_i}{\tau_i}
\label{spin}
\end{equation}

\begin{figure}[t]
\includegraphics[width=250pt]{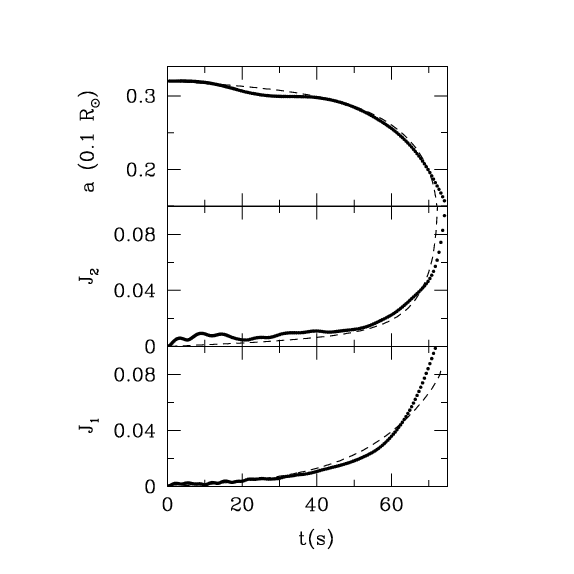}
\caption{A comparison of  the SPH values for the  orbital distance $a$
         and  for  the spin  angular  moments  of  the donor  and  the
         accretor stars  for the  $0.6+0.8\, M_{\sun}$ case  and those
         obtained using a simplified  model. The sping angular moments
         are expressed in units of 10$^{51}$ g cm$^2$/s.}
\label{comparison}
\end{figure}

\noindent  can be  integrated. In  doing so  it has  to be  taken into
account  the  logical limitations  of  the  theoretical approach.   In
particular, the SPH  results show that mass transfer  is not perfectly
conservative, although  this assumption is  fairly good ---  see Table
\ref{tab-hydro}. Moreover,  stars are not point-like  masses and, most
importantly,  we  have not  adopted  in  our  analysis an  equilibrium
mass-radius  relationship.   These   assumptions  may  produce  marked
differences between the SPH  and the theoretical results.  Perhaps the
most critical assumption in determining the evolution of the system is
the assumption that an equilibrium mass-radius relation holds for both
members  of the  binary  system.  In  fact,  at the  beginning of  the
mass-transfer episode  stars are not in  equilibrium. Consequently, we
have  adopted a  different approach.   In particular  when integrating
Eqs. (\ref{spin}) we have computed for each time step the {\sl actual}
moment of  inertia of each  star.  In particular,  in the case  of the
primary white  dwarf for each computed  model we look  for location of
the region with maximum temperature (see Fig.  \ref{snapshots-T}).  We
then compute  the mass  interior to this  shell and  the corresponding
moment of inertia.  For the case  of the donor white dwarf we look for
the region which still has an approximate spherical symmetry (see Fig.
\ref{snapshots-rho}) and we follow  the same procedure adopted for the
accretor.

In  figure \ref{comparison} we  show a  comparison of  the theoretical
results --- shown  as a dashed line --- and the  SPH results --- shown
as dots  --- for the time  evolution of the orbital  separation and of
the  spin angular  moment  of  the accretor  ($J_1$)  and donor  stars
($J_2$).  The  three adjustable parameters adopted  in the theoretical
calculations are,  respectively, $\alpha=145\, M_{\sun}\, R_{\sun}^3\,
/{\rm yr}$, $\tau_1^0=3.50\times 10^{4}$ yr and $\tau_2^0=4.75 10^{-8}$
yr.   As can  be  seen, during  the  first phases  of  the merger  the
agreement is excellent.  Note however that we can only compare the SPH
results with  the theoretical  expectations while the  secondary still
preserves partially  its initial  shape. This is  why we only  show in
Fig.  \ref{comparison}  a reduced time interval,  corresponding to the
first five panels in Figs.  \ref{snapshots-rho} and \ref{snapshots-T}.
For times  longer than  $\sim 70$ s,  the secondary  rapidly dissolves
and,  hence, the approach  followed here  is no  longer valid.   It is
worth  realizing that  $\tau_1^0\gg  \tau_2^0$.  This  means that  the
synchronization timescale of  the primary is much larger  than that of
the secondary.   Accordingly, during  this phase of  the mass-transfer
episode the  donor rapidly synchronizes whereas the  primary does not.
Consequently, orbital angular moment  is transferred from the orbit to
the donor in a short  timescale, thus reducing the orbital separation.
This, in turn,  increases the mass-transfer rate and  the final result
is that the  secondary is rapidly disrupted.  Since  the total angular
momentum  is conserved the  material transferred  to the  primary must
rotate rapidly, thus  producing the characteristic rotational profiles
shown in the left panels of Fig. \ref{2Dvel}.  In summary, the results
of  the hydrodynamic calculations  can be  accurately reproduced  by a
simple model once  all the weaknesses of the  theoretical approach are
correctly taken into account.

\end{subsection}

\begin{subsection}{Gravitational wave radiation}

Gravitational wave  radiation from  Galactic close white  dwarf binary
systems is expected to be  the dominant contribution to the background
noise in the low frequency region, which ranges from $\sim 10^{-3}$ up
to $\sim  10^{-2}$~Hz (Bender et  al.  1998).  Moreover,  since during
the merging process a sizeable  amount of mass is transferred from the
donor star  to the primary  at considerable speeds,  the gravitational
wave  signal is expected  to be  detectable by  LISA (Guerrero  et al.
2004;  Lor\'en--Aguilar  et  al.   2005).   It is  thus  important  to
characterize  which would be  the gravitational  wave emission  of the
white  dwarf mergers  studied here  and to  assess the  feasibility of
dectecting them.

To  compute   the  gravitational  wave   pattern  we  proceed   as  in
Lor\'en--Aguilar et al.  (2005).  In particular, we use the weak-field
quadrupole approximation (Misner et al. 1973):

\begin{equation}
h^{\rm TT}_{jk} (t,\vec{x}) = \frac{2G}{c^4d} \frac{\partial^2 Q^{\rm TT}
_{jk} (t-R)}{\partial t^2}
\label{h}
\end{equation}

\noindent where $t-R=t-d/c$ is the  retarded time, $d$ is the distance
to  the  observer,  and  $Q^{\rm  TT}_{jk} (t-R)$  is  the  quadrupole
moment of the mass distribution, wich is given by

\begin{equation}
\ddot{Q}^{\rm TT}_{jk} (t-R) = \int  \rho(\vec{x},t-R)(x^jx^k-\frac{1}{3}x^2
\delta_{jk})d^3x
\label{Q}
\end{equation}

\noindent To calculate the  quadrupole moment of the mass distribution
using SPH  particles, Eq.~(\ref{Q})  must be discretized  according to
the following expression

\begin{eqnarray} 
\ddot Q^{\rm TT}_{jk}(t-R)&\approx&  P_{ijkl}(\vec{N})\sum^{n}_{p=1} 
m(p) \big\lbrack  2\vec{v}^k(p)\vec{v}^l(p)\cr
&+&\vec{x}^k(p)\vec{a}^l(p)+\vec{x}^l(p)\vec{a}^k(p) \big\rbrack
\label{mom_discretized}
\end{eqnarray}

\noindent where 

\begin{eqnarray} 
P_{ijkl}(\vec{N})& \equiv & (\delta_{ij}-N_iN_k)(\delta_{jl}-N_jN_l)\cr
&-&\frac{1}{2}(\delta_{ij}-N_iN_j)(\delta_{kl}-N_kN_l)
\label{pol_tensor}
\end{eqnarray}

\noindent  is the  transverse-traceless projection  operator  onto the
plane orthogonal to the  outgoing wave direction, $\vec{N}$, $m(p)$ is
the  mass of  each SPH  particle, and  $\vec{x}(p)$,  $\vec{v}(p)$ and
$\vec{a}(p)$   are,   respectively,   its   position,   velocity   and
acceleration.

\begin{figure}[t]
\includegraphics[width=250pt]{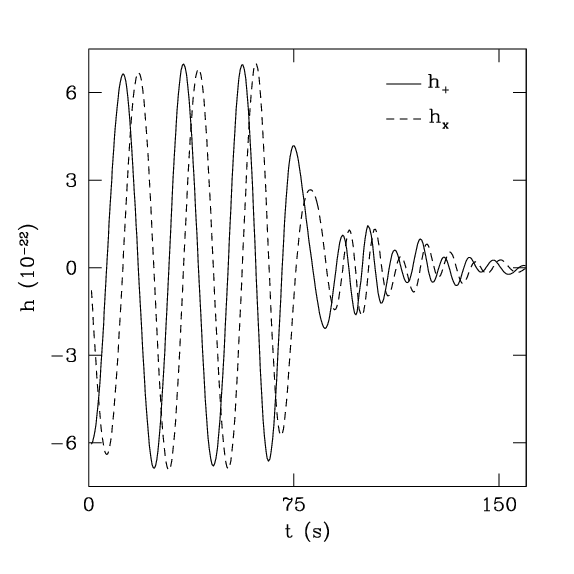}
\caption{Gravitational wave  emission from the merger  of a $0.6+0.8\,
        M_{\sun}$ close white  dwarf binary system.  The dimensionless
        strains  $h_{+}$ and  $h_{\times}$  are measured  in units  of
        $10^{-22}$. The source has been assumed to be at a distance of
        10~kpc.}
\label{GWR}
\end{figure}

Using this prescription, the  corresponding strains for the $0.6+0.8\,
M_{\sun}$,  which  is  a   representative  case,  are  shown  in  Fig.
\ref{GWR}.   As  shown  in  this figure  the  gravitational  waveforms
rapidly vanish  in a couple  of orbital periods and  the gravitational
wave emission during the coalescence  phase does not have a noticeable
large peak. Hence, the gravitational wave emission is dominated by the
chirping phase, in agreement  with the findings of Lor\'en--Aguilar et
al.  (2005).  Moreover, the  gravitational waveforms obtained here are
very similar to those computed  by Lor\'en--Aguilar et al. (2005) and,
thus, do  not depend  appreciably on the  number of particles  used to
calculate them.  This stems from the fact that most of the emission of
gravitational waves comes from the  regions in which the SPH particles
change  appreciably  their velocities,  and  these  regions were  well
resolved  in  both  sets   of  simulations.   Larger  order  terms  of
gravitational wave  emission could be  included in the  calculation of
the strains.  These terms  include the current-quadrupole and the mass
octupole.  It has been shown (Schutz \& Ricci 2001) that for the first
of these  to be relevant an oscillating  angular momentum distribution
with  a dipole  moment  along  the angular  momentum  axis is  needed.
Consequently,  in our calculations  only the  mass octupole  should be
considered in the best of the cases.  Within this approximation a term
of  the order  of  $v/c\sim 10^{-3}$  would  be added  to the  derived
strains.  We  have performed a post-processing of  our simulations and
we have  found that the  octupole emission is  of the order  of $h\sim
10^{-24}$, just  above the numerical noise ($h\sim  10^{-25}$), but in
any case  totally negligible.  Since the gravitational  wave signal is
dominated by  that of the inspiralling  phase, in order  to assess the
feasibility of detecting it using gravitational wave detectors we have
assumed that  the orbital separation of  the binary system  is that of
the last  stable orbit.   Furthermore, we have  also assumed  that the
integration time of LISA will be one year.  It is then straightforward
to demonstrate  that during  this time interval  the variation  of the
orbital separation  is negligible.  With these  assumptions the double
white dwarf binary system  basically radiates a monochromatic wave and
it is easy to assess  the feasibility of detecting the signal produced
by the coalescence of close  binary white dwarf systems.  This is done
in  figure \ref{LISA}  where we  show  the strength  of the  resulting
signals and we compare them with the spectral distribution of noise of
LISA, when a  distance of 10 kpc  is adopted.  As can be  seen in this
figure all the  systems are well inside the  detectability region and,
hence, LISA should be able to distinguish  them  from  Galactic noise.
However, the typical  rate of white dwarf mergers  is rather small, of
the  order of $\sim  9\times 10^{-3}$  yr$^{-1}$ (Nelemans  2003) and,
consequently, although there is an uncertainty of a factor of 5 in the
rate of white  dwarf mergers, the expected detection  rate is small as
well.

\begin{figure}[t]
\includegraphics[width=250pt]{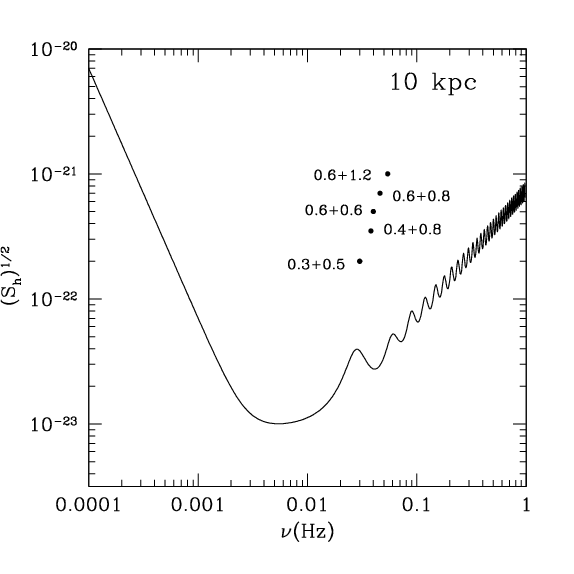}
\caption{A comparison of the signal  produced by the close white dwarf
        binary systems  studied  here  when, a  distance  of 10~kpc is 
	adopted, with the spectral distribution of noise  of LISA. The 
	spectral  distribution of   noise  of  LISA  is for a one year  
	integration  period.  We  have adopted a signal-to-noise ratio 
	$\eta=5$.}
\label{LISA}
\end{figure}

\end{subsection}

\begin{subsection}{Fallback luminosities}

Another potential observational signature  of the mergers studied here
is the emission  of high-energy photons from the  fallback material in
the aftermath of  the coalescence itself.  We have  already shown that
as a  result of the  merger of two  white dwarfs of  different masses,
most of the SPH particles  of the disrupted secondary form a keplerian
disk.  These  SPH particles have circularized orbits.   However, as it
occurs in  the coalescence of  double neutron stars, some  material of
the  secondary is  found to  be in  highly eccentric  orbits  as well.
After  some time,  this material  will most  likely interact  with the
recently formed  disk.  As discussed  in Rosswog (2007)  the timescale
for  this is  not  set by  viscous  dissipation but,  instead, by  the
distribution of eccentrities.  We follow closely the model proposed by
Rosswog (2007)  and calculate  the accretion luminosity  obtained from
the interaction of the  stellar material with high eccentricities with
the newly  formed disk  by assuming that  the kinetic energy  of these
particles is dissipated within the radius of the debris disk.

\begin{figure}[t]
\includegraphics[width=250pt]{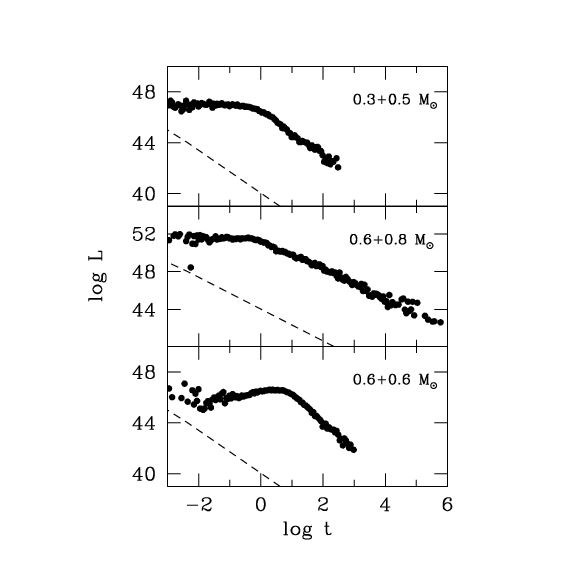}
\caption{Fallback accretion  luminosity for our  three fiducial double
         white dwarf mergers.  The  units of time are seconds, whereas
         those of  luminosities are erg/s. A straight  line with slope
         5/3 is shown for the sake of comparison.}
\label{lumin}
\end{figure}

In figure \ref{lumin} we have  plotted the accretion luminosities as a
function of time for our three fiducial cases. We emphasize that these
luminosities  have been  computed assuming  that the  highly eccentric
particles loose all its kinetic energy when interacting with the disk,
for  which  we  adopt the  radius  obtained  by  the  end of  our  SPH
simulations,   which  are  those   shown  in   Table  \ref{tab-hydro}.
Moreover, only a fraction of this  energy will be released in the form
of high-energy  photons. Thus, the  results shown in  Fig. \ref{lumin}
can be regarded  as an {\sl upper limit} for  the actual luminosity of
high-energy photons.  Note that  although the luminosities are smaller
than those typically  obtained for the merger of  double neutron stars
--- which are typically of the order of $\sim 10^{52}$ erg/s --- white
dwarfs  mergers  predict  a  very similar  time  dependence  ($\propto
t^{5/3}$).

This  is an  important result  because it  shows that  observations of
high-energy  photons  can help  in  detecting  the gravitational  wave
signal  radiated by  these systems.   In  fact, the  detection of  the
gravitational waves arising from the merger of white dwarfs is a tough
task because, as previously explained,  the signal is dominated by the
inspiralling  phase and  the waveforms  do not  have a  prominent peak
before the ringdown phase. Thus  a combined strategy in which optical,
UV,  X-ray and  gravitational wave  detectors are  used could  be very
useful. 

\end{subsection}

\begin{subsection}{Long-term evolution}

We have  already shown that non-explosive nuclear burning  takes place
during the merging phase. However, this does not necessarily mean that
such an explosion could not take  place due to mass accretion from the
disk at late  times.  If mass acccretion occurs  at rates smaller than
$10^{-6}\, M_{\sun} \, {\rm yr}^{-1}$ then, central carbon ignition is
possible and a SNIa is the  most probable outcome.  On the other hand,
if the  accretion rates  are larger than  this value,  then off-center
carbon ignition is the most probable outcome, giving rise to an inward
propagating  burning flame  and an  ONe white  dwarf is  likely  to be
formed (Nomoto \&  Iben 1986; Garc\'\i a--Berro \&  Iben 1994; Ritossa
et  al.   1999)   wich  might  eventually  form  a   neutron  star  by
accretion-induced  collapse (Saio  \& Nomoto  1985; Woosley  \& Weaver
1986; Guti\'errez et al. 1996, 2005).  However, once the disk has been
formed,  angular  momentum  viscous   transfer  is  relevant  and  the
hydrodynamical   timescale   of   the   disk   becomes   very   large.
Consequenlty, the subsequent evolution  of the disk cannot be followed
using a SPH  code.  However, some estimates of  the accretion rate can
still be done by considering the typical viscous transport timescales.

The typical viscous transport timescale is (Mochkovitch \& Livio 1989,
1990)

\begin{equation}
\tau_{\rm visc} = \left(\frac{1}{T}\frac{dT}{dt}\right)^{-1}
\label{ML}
\end{equation}

\noindent where $T$ is the  rotational  kinetic energy and

\begin{equation}
\frac{dT}{dt} = -\int \left(\frac{\partial\Omega}{\partial r} \right)^2
r^2 \eta(r) d^3r
\end{equation}

\noindent is its  rate of change.  In this  expression $\Omega$ is the
angular velocity, $r$ is  the radial cylindrical coordinate and $\eta$
is  the (physical) viscosity  parameter, that  depends on  the viscous
mechanism.   If the  disk  is laminar  and  the viscosity  is that  of
degenerate electrons then $\eta = 2.0\times 10^{-5} \rho^{5/3} \, {\rm
g/cm\;  s}$  (Durisen 1973,  Itoh  et  al.  1987) and  the  associated
accretion rates can be obtained taking into account that

\begin{equation}
\dot T \sim \frac{GM_{*}\dot{M}}{R_{*}}
\end{equation}

\begin{figure}[t]
\includegraphics[width=250pt]{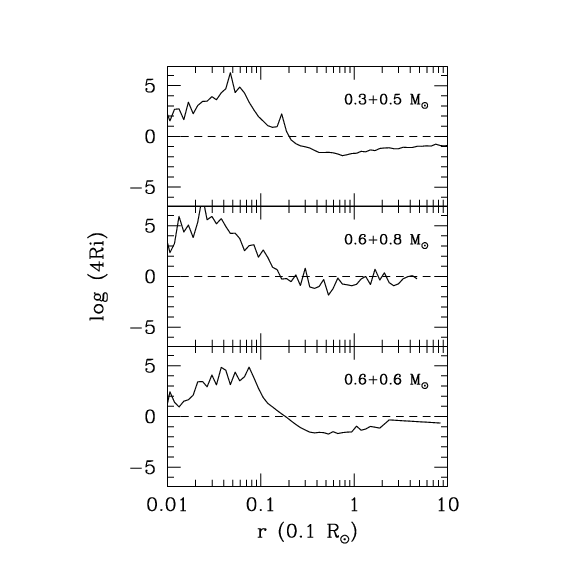}
\caption{Richardson  number  as  a  function  of  the  distance.  When
         Ri$>1/4$ the  disk  is turbulent. The  horizontal dashed line 
         corresponds to Ri=1/4.}
\label{Ri}
\end{figure}

\noindent where $M_*$ and $R_*$ are the mass and radius of the central
object. If, instead, the disk is turbulent the classical approximation
of Shakura \& Sunyaev (1973)  is valid.  Within this approximation the
viscous timescale is given by

\begin{equation}
\tau_{\rm visc}=\alpha^{-1} \left(\frac{R_{\rm disk}}{H}\right) 
\left(\frac{R_{\rm disk}}{c_{\rm s}}\right)
\end{equation}

\noindent where  $\alpha\sim 0.1$  is the standard  viscosity, $c_{\rm
s}$ is the  sound speed, $R_{\rm disk}$ is the radius  of the disk and
$H$ is the  disk half-thickness.  Both the radius of  the disk and its
half-thickness are listed in Table \ref{tab-hydro} for each one of the
simulations presented here.  The accretion rate is then given by

\begin{equation}
\dot M\simeq \frac{M_{\rm disk}}{\tau_{\rm visc}}
\label{SS}
\end{equation}

In  order to  check if  the  disk is  turbulent we  have computed  the
Richardson number

\begin{equation}
{\rm Ri} = \frac{\left(\frac{g^{\rm eff}}{c_{\rm s}}\right)
\left(1-\frac{\gamma}{\Gamma}\right)}
{\left( r\frac{d\Omega}{dr} \right)}
\end{equation}

In this expression $g^{\rm eff}$ is the effective gravity, that is the
real gravity minus the  centrifugal force, $\gamma$ is the logarithmic
derivative of  the pressure with  respect to the density,  $\Gamma$ is
the  adiabatic index  and the  rest of  the symbols  have  their usual
meaning.   We  have  chosen  $\gamma=1.4$ and  $\Gamma=5/3$.   If  the
Richardson number is smaller  than 1/4 stability against turbulence is
guaranteed.  If this is not the  case the disk may be turbulent, since
this is only a necessary  condition, but Ri$>1/4$ is a good indication
for turbulence  to occur.  In  figure~\ref{Ri} we show  the Richardson
number as a  function of the radial coordinate  for the three fiducial
cases described  here.  As  can be seen  in this figure  the condition
Ri$>1/4$  is satisfied in  the innermost  regions of  the disk,  up to
distances $\sim  0.2\, R_{\sun}$.  Thus, the innermost  regions of the
disk are  potentially turbulent  and it is  likely that  the accretion
rate  is that given  by Eq.~(\ref{SS}).   Nevertheless, Eq.~(\ref{ML})
provides a safe lower limit for the typical transport timescale of the
disk.  The  total rotational kinetic luminosity  dissipated, $\dot T$,
and the corresponding timescales using  this approach are given in the
left section of  table \ref{mdot}.  For the case  in which the laminar
viscosity is  used the resulting accretion  rates turn out  to be $\la
10^{-12}\,  M_{\sun}/{\rm   yr}$,  and  consequently   central  carbon
ignition leading to a SNIa is possible.  When the classical Shakura \&
Sunyaev (1973) expression is adopted  the accretion rates are shown in
the right section of table \ref{mdot}.  As can be seen these accretion
rates  turn  out  to  be  very  large.   There  are  experimental  and
theoretical reasons  to suspect  that the central  object will  not be
able to accrete material from  the surrounding disk at these very high
accretion rates.   From a theoretical  perspective it is  rather clear
that these accretion rates exceed  the Eddington limit, which is order
of  $10^{-5}\,M_{\sun}\,  {\rm yr}^{-1}$.   Additionally,  and from  a
experimental point of view, there is growing evidence (Ji et al. 2006)
that  hydrodynamic   turbulence  cannot  transport   angular  momentum
effectively  in  astrophysical  disks,  even at  very  large  Reynolds
numbers, leaving as the only possible way to lose angular momentum the
magnetorotational instability.

\begin{table}[t]
\centering
\begin{tabular}{cccccc}
\hline
\hline
\noalign{\smallskip}
\multicolumn{1}{}{} &
\multicolumn{3}{c}{Laminar} &
\multicolumn{2}{c}{Turbulent} \\
\noalign{\smallskip}
\hline
\noalign{\smallskip}
 Run        & $\tau$  & $\dot T$ & $\dot M$  & $\tau$ & $\dot M$ \\
\noalign{\smallskip}
\hline
\noalign{\smallskip}
0.3+0.5  & $3\times10^{11}$ & $1\times10^{29}$ & $2\times10^{-14}$ & $7.6\times10^{-4}$ & 43  \\
0.4+0.8  & $1\times10^{11}$ & $5\times10^{29}$ & $8\times10^{-14}$ & $1.1\times10^{-2}$ & 63  \\
0.6+0.6  & $1\times10^{11}$ & $6\times10^{29}$ & $1\times10^{-13}$ & $2.0\times10^{-4}$ & 560 \\
0.6+0.8  & $5\times10^{10}$ & $3\times10^{30}$ & $4\times10^{-13}$ & $1.2\times10^{-2}$ & 62  \\
0.6+1.2  & $3\times10^{8}$  & $7\times10^{32}$ & $8\times10^{-11}$ & $1.0\times10^{-2}$ & 75  \\
\noalign{\smallskip}
\hline
\hline
\end{tabular}
\caption{Typical  viscous timescales  (in  years), dissipated  kinetic
         energies  (in  erg/s),  and  accretion rates  (in  $M_{\sun}$
         yr$^{-1}$)  for the case  in which  the laminar  viscosity is
         used.  Columns  5  and  6  show  the   corresponding  viscous 
         timescales  and accretion  rates obtained when the  classical 
         Shakura \& Sunyaev (1973) expression is employed.}
\label{mdot}
\end{table}

Yoon  et al. (2007)  have systematically  explored the  conditions for
avoiding  off-center  carbon  ignition  in the  merged  configurations
resulting from the  coalescence of two CO white  dwarfs. They computed
the  evolution  of  the  central  remnant  of  the  coalescence  of  a
$0.9+0.6\, M_{\sun}$ binary white  dwarf, adopting a realistic initial
model,  which includes  the differentially  rotating primary,  the hot
envelope  we  also  find  in  our simulations  and  the  centrifugally
supported accretion disk.  Our final configurations resemble very much
those found  by these authors and, consequently,  the same conclusions
obtained in  this paper hold. In  particular, since in  our models the
maximum temperature  of the  hot envelope is  smaller than  the carbon
ignition temperature  and the  mass-accretion rate from  the keplerian
disk  is  possibly  smaller  than  the  critical  one  ($\dot  M\simeq
5.0\times 10^{-6}\, M_{\sun}\, {\rm  yr}^{-1}$) it is probable that at
least some  of the merged  configurations with a total  mass exceeding
Chandrasekhar's  mass may be  considered good  candidates for  Type Ia
supernovae progenitors.

\end{subsection}


\section{Conclusions}

We   have   performed   several  high-resolution   Smoothed   Particle
Hydrodynamics simulations of coalescing white dwarfs.  We have done so
for  a  broad  range  of  masses  and  chemical  compositions  of  the
coalescing white dwarfs,  which includes He, CO and  ONe white dwarfs.
Such a  parametric study using  a large number of  particles ($4\times
10^5$ SPH particles) had never been done before. Previous works on the
subject used a  considerably smaller number --- by a  factor of 10 ---
of SPH particles  (Guerrero et al. 2004), or did  not explore the full
range of  masses and  chemical compositions of  interest (Yoon  et al.
2007) --- only one  merger was computed, that  of $0.9+0.6\, M_{\sun}$
double white dwarf.  In addition, we have included a refined treatment
of  the  artificial  viscosity.    In  particular,  we  have  used  an
artificial viscosity  formulation that is  oriented at Riemann-solvers
(Monaghan  1997) together with  an additional  switch to  suppress the
excess  of   viscosity  (Balsara   1995).  With  this   treatment  the
dissipative terms  are only applied in  those regions of  the fluid in
which  they are  really necessary  to  resolve a  shock. This  refined
treatment of  the artificial viscosity overcomes some  of the problems
found in our previous simulations (Guerrero et al. 2004).

In all cases,  the merged configuration consists of  a compact central
object  surrounded  by  a  hot  corona with  spheroidal  shape  and  a
self-gravitating keplerian disk around it.  For the cases in which two
white dwarfs of  different masses are involved the  resulting disk can
be considered  as a  thin disk, whereas  for the  $0.6+0.6\, M_{\sun}$
case we have found that  the resulting final configuration resembles a
rotating ellipsoid around  the central object with a  much more modest
disk.  The  peak temperatures attained during the  merging process are
smaller than  those found by Guerrero  et al. (2004) and  in line with
that  found by  Yoon  et al.   (2007)  for the  case  of a  $0.9+0.6\,
M_{\sun}$ merger.   We also confirm  the results obtained  in previous
works (Guerrero et  al.  2004: Lor\'en--Aguilar et al.   2005; Yoon et
al.  2007) and we find that  only when one of the merging white dwarfs
is a He  white dwarf nuclear reactions are  relevant.  However none of
the cases studied  here show an explosive behavior  during the merging
phase.   Furthermore,  no essential  differences  are  found when  the
chemical abundances obtained here using an enhanced spatial resolution
and  a refined  prescription for  the artificial  viscosity  and those
obtained  in  previous works  (Garc\'\i  a--Berro  et  al.  2007)  are
compared.  The  chemical  composition  of  the  disks  formed  by  the
coalescence  of  a  He white  dwarf  with  CO  white dwarfs  shows  an
enrichment in heavy elements like Ca,  Mg, S, Si and Fe and constitute
a natural environment where planets  and asteroids can be formed. This
could  explain the  anomalous abundances  of metals  found  in several
hydrogen-rich white dwarfs with dusty disks around them and which have
been attributed  to the impact of  asteroids (Jura 2003),  since it is
quite unlikely that such asteroids could survive the red giant phase.

We have  also compared the results of  our hydrodynamical calculations
with  the theoretical expectations  and we  have found  a satisfactory
agreement  when   the  synchronization  timescale   of  the  disrupted
secondary is much shorter than that  of the primary.  In this case the
rate of change of the  orbital distance and the corresponding spins of
both the  donor star and of  the accretor are reproduced  with a large
degree  of  accuracy. We  have  shown as  well  that  the emission  of
gravitational waves from these kind  of systems is strong enough to be
obervable by LISA, and that  the corresponding waveforms do not depend
appreciably   on  the  resolution   employed  in   the  hydrodynamical
calculations and, thus, that these waveforms are robust.

We have computed as well  the possible emission of high-energy photons
produced in the aftermath of  the merger. This high-energy emission is
a consequence of the interaction of the material with highly eccentric
orbits --- which is produced  during the first and most violent phases
of  the merger  --- with  the resulting  disk ---  which is  formed by
particles  with circularized  orbits ---  and we  have found  that the
typical luminosities  are of the order $\sim  10^{49}$ erg/s, although
the  precise value of  the peak  luminosity depends  very much  on the
masses of  the coalescing  white dwarfs.  The  time dependence  of the
high-energy emission is $\propto  t^{5/3}$, a behavior also present in
the coalescence  of double neutron  stars and neutron stars  and black
holes (Rosswog  2007).  The detection of this  high-energy burst would
eventually  help  in   detecting  the  gravitational  waves  suposedly
radiated during the merger.

With respect to  the long-term evolution of the  mergers we have found
that all the disks product of  the coalescence of two white dwarfs are
potentially turbulent.  This result  implies that very large accretion
rates from the  disk onto the primary are  expected. Despite our crude
approximations,  it is thus  quite likely  that these  accretion rates
would  lead to  an off-center  carbon ignition,  although an  in depth
study  remains to be  done.  However,  since our  final configurations
resemble very  much those found by Yoon  et al. (2007), it  is as well
probable  that at  least  some  of our  merged  configurations may  be
considered good  Type Ia supernovae candidate  progenitors. A detailed
calculation  of the  evolution of  the resulting  disks,  including an
accurate description of the  mechanisms of angular momentum transport,
must  therefore be  done. Unfortunately  this task  is far  beyond the
current possibilities of SPH techniques.


\begin{acknowledgements}
Part   of    this   work   was    supported   by   the    MEC   grants
AYA2008--04211--C02--01 and 02, ESP2007--61593, by the  European Union
FEDER funds and by the AGAUR.
\end{acknowledgements}


\end{document}